\documentstyle[aasms4]{article}

\begin{document}
\def\ptsec{$''\mskip-7.6mu.\,$}
\def\psec{$^s\mskip-7.6mu.\,$}
\def\arcsec{$^{\prime\prime}$}

\def\Msun{\,{\rm M$_{\odot}$}}
\def\Lsun{\,{\rm L$_{\odot}$}}
\def\arcmin{{$'$}}
\def\ptsec{$''\mskip-7.6mu.\,$}
\newcommand{\mm}{\,${\rm mm}$}
\def\kms{\,{\rm {km~s^{-1}}}}
\newcommand{\degr}{\mbox{\,$^\circ$}}        
\newcommand{\mic}{\mbox{\,${\mu}$m}}

\title{A Submillimeter Study of the Star-Forming Region NGC 7129}

\author{Andreea S. Font\altaffilmark{1} and George F. Mitchell}
\affil{Department of Astronomy and Physics, Saint Mary's University, Halifax, NS, B3H 3C3, Canada}
\altaffiltext{1}{Present address: Department of Physics and Astronomy, University of Victoria, P.O Box 3055, Victoria, B.C., V8W 3P6, Canada.~~afont@noir.phys.uvic.ca}
\author{G\"oran Sandell\altaffilmark{2}}
\affil{USRA, NASA Ames Research Center, MS 144-2,
Moffett Field, CA 94035, U.S.A.}
\altaffiltext{2}{gsandell@mail.arc.nasa.gov}

\begin{abstract}


New molecular ($^{13}$CO J=3$-$2) and dust continuum (450\mic\ and 850\mic) maps of the NGC7129 star forming region are presented. The maps include the Herbig Ae/Be star LkH$\alpha$ 234, the far-infrared source NGC 7129 FIRS2 and several other pre-stellar sources embedded within the molecular ridge. The data are complemented with C$^{18}$O J=3$-$2 spectra at several positions within the mapped region. 
Both the submillimeter and the $^{13}$CO emission show a similar morphology, displaying a sharp boundary towards the cavity. The submillimiter maps also reveal a second source, SMM\,2, which is not clearly seen in any earlier data set. This is either a pre-stellar core or possibly a protostar. Also, the highest continuum peak emission is identified with the deeply embedded source IRS\,6 a few arcseconds away from LkH$\alpha$ 234. These new 450 and 850\mic\ observations are combined with previous continuum observations of the three compact far-infrared sources in the field, in order to make fits to the spectral energy distributions and to obtain the source sizes, dust temperatures, luminosities, and masses. For nine positions where both $^{13}$CO and C$^{18}$O spectra are available, gas masses have been obtained and compared with masses derived from the continuum fluxes. The masses are found to be consistent, implying little or no CO depletion onto grains. The dust emissivity index is found to be low towards the dense compact sources, $\beta$ $\sim$1 $-$ 1.6, and high, $\beta$ $\sim$ 2.0, in the surrounding cloud.

\end{abstract}


\section{Introduction}

NGC 7129 is a reflection nebula seen against a molecular cloud and
estimated  by Shevchenko {\it et al.} (\cite{Shev89}) to be at a
distance of 1.25 kpc.  There are many signs in this region which
support the idea of triggered and possibly ongoing star formation: the
association of a very young cluster formed by the B3-type stars BD +65
1637 and BD +65 1638, and the B5/7-type star LkH$\alpha$ 234, several
embedded infrared sources (Harvey {\it et al.} \cite{Harvey84};
Weintraub {\it et al.} \cite{Wein94}; Cabrit {\it et al.}
\cite{Cabrit97}) as well as reflection nebulae and Herbig-Haro (HH)
objects (Hartigan \& Lada \cite{Hart85}; Miranda {\it et al.}
\cite{Miranda94}). There are also several molecular outflows in the
region: one associated with an optical jet near LkH$\alpha$ 234 (Ray
{\it et al.}, \cite{Ray90}; Edwards \& Snell \cite{Edwards83}),
another bipolar outflow near the far-infrared source NGC7129 FIRS 2 and
one may be driven by the T Tauri star V350 Cep (Hartigan \& Lada
\cite{Hart85}; Goodrich \cite{Goodrich86}; Miranda {\it et al.}
\cite{Miranda94}).

A molecular cavity is revealed in both molecular and infrared maps
(Bechis {\it et al.} \cite{Bechis78}; Bertout \cite{Bertout87}), but
its origin is still unknown. The location of the two bright stars, BD
+65 1637 and BD +65 1638, inside the cavity, suggests three mechanisms by which the
molecular cavity could have been created: {\it (i)} through the
advancement of a photon-dominated region (PDR) around those stars, in
which the molecular gas was dissociated by their intense UV radiation,
{\it (ii)} by an expanding shell associated with stellar mass loss, or
{\it (iii)} by radiation pressure acting on grains, together with
gas-grain drag (Bechis {\it et al.} \cite{Bechis78}). The two mid-B
type stars in the cavity, BD +65 1637 and BD +65 1638, are also the 
oldest in the cluster. Therefore stellar mass loss may have been very 
effective at
earlier times and the winds from those two stars may have triggered the
formation of other bright stars in NGC 7129, including LkH$\alpha$ 234
and SVS 13.  As for what mechanism is dominating at present time, the
question remains open. PDR models can marginally explain the line
fluxes of [OI] and [CII] emission in the region, although the C-type
shock models cannot be ruled out (Lorenzetti {\it et al.}
\cite{Loren99}; Giannini {\it et al.} \cite{Giannini99}).

This paper employs new, high resolution submillimeter observations of the molecular ridge in dust continuum emission, $^{13}$CO line emission, and C$^{18}$O line emission. One goal of this study is to improve the present picture of the structure of the ridge and the ridge-cavity boundary. Another goal is to obtain the physical properties of the compact far-infrared sources in the ridge. A third aim is to obtain the gas/dust ratio, thus determining whether CO isotopomers are a good surrogate for molecular hydrogen, or whether CO has substantially frozen out onto dust grains. The final goal is to search for  variations in the dust opacity with position through the field. The mapped region includes the dense molecular cloud, a photodissociation region, and a powerful outflow. We might expect processes such as grain coagulation and mantle evaporation to have a strong spatial dependency here and for this to be reflected in spatial variations of the dust emissivity index, $\beta$.   

\section{Observations and data reduction}

\subsection {$^{13}CO$ and C$^{18}$O observations}

Observations of $^{13}$CO and C$^{18}$O J=3$-$2 were made with the
James Clerk Maxwell Telescope, on Mauna Kea, Hawaii, in August 1994 and 
November 1997 respectively, with the receiver B3i and the Digital Autocorrelation Spectrometer (DAS). The observations were made by position switching, with the reference position 10$^{\prime}$ East of LkH$\alpha$ 234, which  was determined to be free of significant CO emission.  The main beam efficiency ($\eta_{bm}$) was 0.58 and the half-power beam (HPBW) was $\sim$
14\arcsec\ for both $^{13}$CO and C$^{18}$O observations.


A grid map with the spacing of 7\ptsec5 x 7\ptsec5 was taken in the $^{13}$CO
J=3$-$2 (330.58 GHz) rotational transition. The map was centered near the
star LkH$\alpha$ 234\footnote{The astrometric position for LkH$\alpha$ 234 given by Clements \& Argyle (\cite{Clements84}) is $\alpha$(2000) = 21$^{h}$ 43$^{m}$ 06$^{s}$.802 and $\delta(2000)$ = 66$^{\circ}$ 06$^{\prime}$ 54\ptsec5.}, at coordinates $\alpha(2000)$ = 21$^{h}$ 43$^{m}$ 06$^{s}$.08 and $\delta(2000)$ = 66$^{\circ}$ 06$^{\prime}$ 56\ptsec29, and extending from $-$100\arcsec\ to +100\arcsec\ in $\alpha$ and from $-$150\arcsec\ to 75\arcsec\ in $\delta$ offsets. The instrumental velocity resolution was 0.28 km s$^{-1}$. C$^{18}$O J=3$-$2 (329.33 GHz) spectra were taken at nine positions around the star LkH$\alpha$ 234 and near the molecular ridge. Figure~1 shows the $^{13}$CO integrated intensity ($\int T_{B}dv$) contour map superimposed on a K$^{\prime}$ image of the NGC 7129 region from Hodapp (\cite{Hodapp94}). The nine C$^{18}$O offsets with respect to the star LkH$\alpha$ 234 are shown with cross symbols.

\subsection{Submillimeter continuum observations}

Two deep 850 $\mu$m and 450 $\mu$m--maps of LkH$\alpha$\,234 and 
the bright-rimmed cloud ridge southwest of LkH$\alpha$\,234 
have been obtained using the Submillimeter Common User Bolometer Array,
SCUBA (Holland {\it et al.} \cite{Holland99}), on the 15\,m James
Clerk Maxwell Telescope, on Mauna Kea, Hawaii, on October 15, 1997. 
The weather conditions were dry and stable with
an 850 $\mu$m zenith optical depth of $\sim$ 0.19.  The two overlapping
fields were obtained in jiggle-map mode with a chop-throw of
150\arcsec\ in Right Ascension with a total integration time of 85
minutes/field. For a description of SCUBA and its observing modes, see
Holland {\it et al.} (\cite{Holland99}).  Unfortunately the maps were
taken without usable calibration and pointing observations and we have
therefore reduced the data by determining the HPBW of the telescope
from observations of Uranus from other nights in September and October
with similar sky conditions. The same nights were also used to
determine gain conversion factors for the 850 $\mu$m and 450 $\mu$m
filters.  The HPBW was found to be $\sim$ 15\ptsec6 $\times$ 13\ptsec5
at 850 $\mu$m  and 9\ptsec1 $\times$ 7\ptsec8 at 450 $\mu$m with the
beam broadened in the chop direction.  The submillimeter position of
LkH$\alpha$\,234 (Table~3), which has an accuracy of $\sim$ 1\arcsec, was taken
from three short-integration maps interlaced with pointing observations
in SCUBA commissioning time during Spring 1997. These maps have not been 
added in the final data set, because they were taken with a
120\arcsec\ Azimuth chop and there is clear evidence that the array was
chopping onto emission. However, if we compare the total flux density
of LkH$\alpha$\,234 (i.e.  background subtracted flux density) it
agrees very well with data taken in October.

We also include a map of the H$_2$O maser and bipolar outflow source
NGC\,7129 FIRS2, which was used as a test source during SCUBA
commissioning time (Sandell \cite{Sandell97}). A composite map at 850
$\mu$m, including all the three overlapping fields is shown in
Figure~2.

The basic data reduction was done with SURF (Jenness \& Lightfoot
\cite{Jenness99}). Even with a 150\arcsec-chop it is more than likely
that we may have chopped onto some emission in both fields, but 
it does not appear to affect the morphology of the
strong dust ridge in which LkH$\alpha$\,234 is embedded. Some
negative emission is detected west of LkH$\alpha$ 234, 
which confirms that there is some emission in the off-position. 
It is therefore likely that emission east
of the ridge is also affected, which is why the 850 and 450\mic\ maps
appear somewhat different. To map a cloud ridge in beam switch mode is
not very advisable, but at the time SCUBA was only available in jiggle
map mode. Although chopping into the cloud cannot be avoided, as long as
the off-source emission is faint and relatively uniform, the morphology 
of the cloud ridge should not be affected. 
However, since the beam size and error lobe contribution
differ between 850\mic\ and 450\mic, this leads to systematic errors
in the integrated intensities (Section ~3.2).

The peak fluxes are uncertain to about 10$\%$ for 850\mic\ and 20$\%$ for 
450\mic.
The final pointing corrected and calibrated maps have an rms noise of
$\sim$ 15 mJy/beam and $\sim$ 120 mJy/beam for 850 and 450\mic,
respectively. These maps were written out as FITS files and read into
the MIRIAD package (Sault, Teuben, \& Wright \cite{Sault95}) for further analysis.

\section{Analysis}

\subsection{$^{13}$CO and C$^{18}$O Emission}

A sharply bounded molecular cavity is seen in $^{13}$CO J=3$-$2
emission westward of LkH$\alpha$ 234, most probably excavated by the
B3-type star BD +65 1637. The lack of molecular emission and the ridge
morphology were also noted in the CO J=1$-$0 and NH$_{3}$ observations
(Bechis {\it et al.} \cite{Bechis78}; Fuente {\it et al.}
\cite{Fuente98}; G\"usten \& Marcaide \cite{Gusten86}). However, our
high resolution (half-beam sampling with a 14\arcsec\ beamsize)
observations resolve the ridge structure in much 
more detail.

There are four $^{13}$CO peaks along the ridge, visible in Figure~1 
as contour peaks. These $^{13}$CO structures are assumed to be physical clumps and named NGC7129 CO1, NGC 7129 CO2, NGC 7129 CO3, and NGC 7129 CO4. 
The positions of these four $^{13}$CO peaks, together with their peak antenna 
temperatures, line widths, and line center velocities for the $^{13}$CO 
emission are given in Table~1. NGC7129 CO1 coincides within errors with LkH$\alpha$ 234 (position 5 in Table~2), NGC 7129 CO4 with position 7 and the other two clumps, NGC7129 CO2 and NGC7129 CO3 are in located close to the positions 6 and 3, respectively.

Perhaps the most striking feature of the $^{13}$CO map is the sharpness
of the transition between the molecular cavity to the west and the
molecular cloud to the east. $^{13}$CO line strengths drop by factors
of 20 or more across the boundary. The narrowness of the transition
region between atomic and molecular gas has implications for the three
dimensional topology of the region. There seem to be three
possibilities: Firstly, the boundary region may be a sheet which
happens to be perpendicular to the line-of-sight. Secondly, the
molecular cavity may be a bowl, which is surrounded by molecular gas.
Thirdly, the molecular cloud may take the form of a rather narrow ridge, 
which is partially surrounded by atomic gas. The first possibility is
improbable, while the second should lead to a broader transition region
(molecular to atomic). We favor the third picture, in which the dense
portion of the molecular cloud consists of a clumpy ridge with a
line-of sight extent comparable to its extent in the plane of the 
sky (roughly 0.5 pc).

Another remarkable feature which can be seen in Figure~1 is the narrow
filament of emission seen in the K$^{\prime}$ image (Hodapp 1994). It extends roughly south from LkH$\alpha$ 234 and then curves to the west, keeping a nearly constant surface brightness. In an unpublished observations using an IR camera with a circularly variable filter on the CFHT, Mitchell and Nadeau found this filament to be also a source of strong emission in the S(1) v= 1$-$0 line of H$_{2}$. The optical filament follows closely the outermost
$^{13}$CO contours, which separate the cavity from the molecular cloud.
Can this close vicinity be explained by standard PDR models? 
The presence of the photodissociation region is a natural explanation for both the sharp molecular boundary and the line-emitting filament. 
The H$_{2}$ S(1) v= 1$-$0 line emission is then due to
fluorescence. The PDR is very likely produced by UV radiation from the
early B star BD+65 1638. However, the detection of all three 
types of emission: optical, CO and H$_{2}$ v= 1$-$0, at the same location, is in contrast with predictions of steady state stationary PDR models (Hollenbach \& Tielens \cite{Hollenbach97}). These models predict a more gradual succession of emission layers as one moves away from the UV ionization front, and where 
the H$_{2}$ v= 1$-$0 transition layer would be separated by about
10\arcsec\ from the molecular CO emission. On the other hand, the observed 
morphology may be explained in the framework of nonstationary PDR
models (Bertoldi \& Draine \cite{Bertoldi96}), which show that the
inhomogeneities in the molecular cloud surface can lead to the merging
of the ionization and the dissociation front.

Because $^{13}$CO is likely to have an appreciable optical depth,
C$^{18}$O 3$-$2 spectra were obtained at several positions. The
positions for which we have C$^{18}$O spectra are marked by the
cross symbols in Figure~1. A comparison between the $^{13}$CO and C$^{18}$O
 spectra at the nine offset positions is shown in Figure~3. Positions
are given as offsets in arcseconds from LkH$\alpha$ 234. The line
intensities are given in antenna temperatures, ${T_{A}}^{*}$, after
corrections for losses due to the Earth atmosphere and the telescope.
For all spectra only the linear baselines were removed, using the SPECX
software package (Padman \cite{Padman92}).

Near LkH$\alpha$ 234, the lines are broadened due to contributions from
the outflow. Position 1 (Table~2) is near the peak of CO emission in the 
redshifted outflow lobe. Both the $^{13}$CO and C$^{18}$O spectra are double 
peaked at this position. The value of the line ratio shows that C$^{18}$O 
is optically thin. The two peaks therefore represent two real kinematic 
components rather than a single line suffering self absorption. The peak 
at $-$10 km s$^{-1}$ is the  ambient gas at this position, while the peak 
at $-$7 km s$^{-1}$ is the postshock gas.

Several line parameters have been calculated at each offset, assuming
local thermodynamical equilibrium (LTE) conditions. A value of 10 was used 
for the abundance ratio ${N_{^{13}CO}}$/${N_{C{^{18}O}}}$, estimated from the partial ratios ${N_{C{^{18}O}}}/N_{H_{2}} \sim $ 2 $\times$ 10$^{-7}$ (Frerking, Langer $\&$ Wilson \cite{Frerking82}) and the standard value  ${N_{^{12}CO}}/{N_{^{13}CO}} \sim $ 56 (Wilson and Rood, \cite{Wilson94}). The intensity ratio, $^{13}$CO/C$^{18}$O, was used to find the optical depth, assuming a common T$_{ex}$ for the two transitions. The value of T$_{ex}$ was computed from the LTE expression relating line strength (given by the radiation temperature), line optical depth $\tau$, the excitation and radiation temperatures:

\begin{equation}
T_{ex} = \frac{h\nu}{k}\frac{1}{\ln[1+(1-e^{-\tau})h\nu/kT_{R}]}.
\end{equation}

In the above relation, the radiation temperature, $T_{R}$, is related to the antenna temperature, ${T^{*}}_{A}$, by:

\begin{equation}
T_{R} = \frac{{T^{*}}_{A}}{\eta_{bm}f},
\end{equation}

\noindent 
where the beam efficiency $\eta_{bm}$ was 0.58 and the filling factor f was assumed to be 1. We will further assume that the gas kinetic temperature is equal to the excitation temperature, but caution that a density greater than 10$^{4}$ cm$^{-3}$ is required for this to be true. The results of the LTE analysis are shown in Table~2. For LkH$\alpha$ 234, the excitation temperature is found to be about 30 K, in agreement with previous values of 20 $-$ 30 K obtained from lower resolution (70\arcsec) CO observations (Bechis {\it et al.} \cite{Bechis78}) and
with the 20 $-$ 25 K obtained from NH$_{3}$ (40\arcsec) observations (G\"usten \& Marcaide \cite{Gusten86}).

At position 6 (Table~2), the gas has an excitation
temperature of $\sim$ 34 K, which is higher than the typical value for a dark
core, but lower than the previous value of 63 K obtained by Mitchell
and Matthews (\cite{Mitch94}) from $^{12}$CO and $^{13}$CO data. These two
temperatures can be reconciled, taking into account that the isotopomers from which they have been derived ($^{12}$CO for 63K and $^{13}$CO for 34 K) have different optical depths and hence probe different regions of the gas. With its lower optical depth, $^{13}$CO will probe deeper into the PDR, where the temperature is lower.

The three last entries in Table~2 represent positions increasing in
distance from the cloud boundary. The excitation temperature is found
to decrease with distance from the boundary, from 38 K through 31 K to
21 K. This decrease is expected as the stellar radiation
is increasingly attenuated with distance into the molecular cloud. At position 1 (Table~2), which is farthest from the cloud boundary, the excitation 
temperature of the ambient gas is lowest, $\sim$ 13 K.

Since the clumps have approximately the same size as our beam and since
the beam does not always include a clump of gas, it is reasonable to
calculate the masses of the gas within the beam (14\arcsec). The
results, including the fractional He abundance and assuming a
distance of 1.25 kpc to the source, are shown in Table~2. The NGC 7129 CO1 
clump which coincides with LkH$\alpha$ 234, contains 10.7 M$_{\odot}$. 
The masses within the beam, at the two offsets adjacent to the clump matching 
the pre-stellar source LkH$\alpha$ 234 SMM\,2 (positions 3 and 6 in Table~2) 
are 12.8 and 7.2 M$_{\odot}$, respectively. However, those values should be 
regarded as upper limits since the temperatures are probably lower, as 
suggested by the more optically thin NH$_{3}$ data. 

These results also depend on the accuracy in the distance determination. The distance estimate of 1kpc to NGC7129, reported by Racine (\cite{Racine68}), has been most commonly used. In our analysis we have adopted the value of 1.25 kpc, determined by Shevchenko {\it et al.} (\cite{Shev89}) on an improved photometry and spectral classification of 35 stars. However, if the PDR is associated with to the Cepheus Bubble, as suggested by {\'{A}}brah{\'{a}}m , Bal{\'{a}}zs \& Kun (\cite{Abraham99}), then the distance to could be as low as 500$-$700 pc, and of course, the derived masses and luminosities would be significantly reduced.

The masses obtained in this analysis will be discussed and compared to masses obtained from the dust emission, in Section 4 below.
 
\subsection{Dust Emission}

In order to compare the 850 and 450\mic\ dust emission with the
molecular line observations, the 450\mic\ images need to be convolved
to the same HPBW as the rest of the observations. However, since the JCMT
telescope is far from perfect at 450\mic, one first needs to remove the 
error beam. A model beam was therefore constructed, by fitting
three Gaussians to representative beam maps of Uranus. The main beam
and the near error lobe were fitted with elliptical Gaussians, while the
far error lobe was fitted with a circularly symmetric beam.  The
relative amplitude of the main beam and the near and far error lobes
were found to be 0.925:0.07:0.005. The HPBW of the near error lobe was
37\arcsec $\times$ 26\arcsec, with the major axis aligned in the chop
direction similarly to the main beam (c.f. Section ~2.2), and the far
error lobe gave a HPBW $\sim$ 120\arcsec. The deconvolution of images was tested with both MAXEN (MIRIAD's maximum entropy task) and CLEAN.  Both
worked reasonably well, but CLEAN is better in preserving flux, and was 
therefore adopted. We also constructed a model beam for 850\mic\ 
and and ran CLEAN on the 850\mic--images as well.  The
deconvolved images were restored with a symmetric Gaussian to a
14\arcsec\ beam. For 450\mic\ we also made maps restored to a HPBW
of 8\arcsec. Figure~4 shows the high resolution maps at 450\mic\ 
and 850\mic.

Even CLEAN, however, is not ideal for recovering smooth extended
emission, which could cause to underestimate the 450 $\mu$m emission
more than at 850\mic, since the 450 $\mu$m--map is restored
from an 8\ptsec5 resolution to 14\arcsec, while the resolution in the
850\mic--maps remains roughly the same.  Emission in the off position
will have an even more severe systematic effect.  Since the error beam
has a much higher amplitude at 450\mic\ than at 850\mic, any
emission in the off position will affect the baselevel of the 450
\mic--map much more than that at 850\mic. This leads to a
systematic underestimate of the flux densities at 450\mic. This 
was tested by adding a constant flux level of 5\% of the peak value at
450\mic\ to the ``raw'' 450\mic--map of LkH$\alpha$\,234. The
baselevel corrected map was run through CLEAN and restored to 14\arcsec.
When the flux densities in the 14\arcsec\ beam are compared for positions
more than 40\arcsec\ away from LkH$\alpha$\,234, it is found that adding
a zero-level of 5\% of the peak flux density increases the flux level
by more than a factor of two compared to the uncorrected map.
Since the base level can easily have an error of 5\%, this means
that the flux densities for faint extended emission are uncertain by
more than a factor of two. In this case the 450\mic\ flux densities
are more than likely to be systematically underestimated.

Two-dimensional Gaussian fits were then used to determine the size and total
flux of the two ``compact'' sources in the field, i.e.
LkH$\alpha$\,234\, SMM\,1 and the protostellar source,
LkH$\alpha$\,234\, SMM\,2, $\sim$ 28\arcsec\ North-West of LkH$\alpha$\,234\, SMM\,1. Here the flux estimates are much more accurate, since the sources are
compact and the emission from the surrounding cloud
ridge can be substracted. LkH$\alpha$\,234\, SMM\,1 emission 
is found to be extended, with a size of $\sim$ 5\arcsec\ (Table~3). 
This agrees well with earlier 800\mic\ and
450\mic\ maps (Sandell \& Weintraub \cite{San_Wei94}) as well as with
1.3\,mm maps by Fuente {\it et al.} (\cite{Fuente98}) and Henning {\it et
al.} (\cite{Henning98}). The source LkH$\alpha$\,234\, SMM\,2 is not
obvious in the 1.3\,mm data of Fuente {\it et al.} (\cite{Fuente98}) nor in the
map published by Henning {\it et al.} (\cite{Henning98}), but both 1.3\,mm maps
show an extension towards SMM\,2 in good agreement with our SCUBA maps. By
using the size derived from our SCUBA maps, we derive an integrated 1.3\,mm 
flux of 0.2 $\pm$ 0.04 Jy for the map published by Fuente {\it et
al.} (\cite{Fuente98}). We are interested to derive the $\beta$-index
(where $\beta$ reflects the change in dust emissivity or, equivalently,
the change in optical depth, with frequency, $\kappa \propto \tau \propto
\nu^{\beta}$) and the total (gas + dust) mass.  For this, we have
fitted the 1.3\,mm data point and the two SCUBA data points with a
simple isothermal model (see e.g.  Sandell \cite{Sandell00}). Since
both IRAS and Kuiper Airborne Observatory (KAO) data 
(Bechis {\it et al.} \cite{Bechis78}; Harvey,
Wilking \& Joy \cite{Harvey84}; Di Francesco {\it et al.} \cite{diFrancesco98}) employ large beam sizes that are likely
to include hot dust from the surrounding reflection nebulosity as well,
we have not included FIR data in our least squares fit, although we
constrained our LkH$\alpha$\,234\, SMM\,1 model to the peak dust temperature
derived by Harvey, Wilking \& Joy (\cite{Harvey84}). The results of the
fit are given in Table~4 and also shown in Figure~5. 
LkH$\alpha$\,234\, SMM\,2 is even less constrained. However, by looking at gas
temperatures, either from our own CO observations or at the temperature 
estimates from the NH$_3$ map by G\"usten \& Marcaide (\cite{Gusten86}),
 we find the gas temperatures to be $\leq$ 30\,K.
Therefore the fit presented in Table~4 was constrained to 30\,K. 
The mass we derive for the dust disk or envelope surrounding LkH$\alpha$\,234\, SMM\,1 (Dent {\it et al}, \cite{Dent89}), is about half of that derived by Henning {\it et al.} ({\cite{Henning98}) after it is adjusted to the same distance. The difference is largely due to the difference in the adopted mass opacity for the dust.

In order to compare the gas mass estimates derived from our $^{13}$CO and
C$^{18}$O data with dust mass estimates, submillimeter flux densities 
have been derived from our submillimeter maps convolved with a 14\arcsec\ beam.
The results are given in Table~5. If one uses the baseline un-adjusted 450\mic\ data and ignores the differences in dust properties and dust temperatures between the two embedded stars by assuming that the dust is at the same temperature as the CO, the $\beta$-values appear unplausibly low. If instead the baseline adjusted 450\mic\ data were to be used one would obtain a $\beta$ of $\sim$ 1.7. However, since we cannot accurately correct for the emission in the 
off position, we prefer to assume a $\beta$-index and only use 
the flux density measurements at 850\mic, which are much less affected by 
emission in the off position. The CO temperatures 
are likely to be dominated by the gas heated by the PDR, and they therefore give only an upper limit to the mass averaged 
gas temperatures. The dust emission, however, is more optically thin and 
the submillimeter dust emission contains also emission from the colder dust 
inside  the dense cloud ridge. 
This regime is more accurately measured by NH$_3$, 
which shows kinetic temperatures of $\sim$ 25 K 
(G\"usten \& Marcaide \cite{Gusten86})
along the cloud ridge. In the second entry for each offset 
lower gas temperatures are assumed, which are typically smaller than the CO 
temperatures. The two compact sources in the ridge have been substracted from the map in order to separate the dust cloud from the disk/envelope emission. However, the total mass values in Table~5 include also the contribution from LkH$\alpha$\,234\, SMM\,1 and LkH$\alpha$\,234\, SMM\,2. 

\subsection{NGC\,7129 FIRS\,2}

NGC\,7129 FIRS\,2 was first discovered by Bechis {\it et al.}
(\cite{Bechis78}) as a cold FIR source associated with a $^{13}$CO
column density peak south of LkH$\alpha$ 234.  It was found to coincide
with an H$_2$O maser (Cesarsky {\it et al.} \cite{Cesarsky78};
Rodr{\'{\i}}guez {\it et al.} \cite{Rodriguez80}; Sandell \& Olofsson
\cite{Sandell81}), suggesting that it is a young protostellar source,
yet the source is invisible in the optical and near-IR. Additional FIR
mapping by Harvey, Wilking \& Joy (\cite {Harvey84}) confirmed the
discovery by Bechis {\it et al.} (\cite{Bechis78}) and detected the
source at both 50 and 100 $\mu$m, but not at 20 $\mu$m or shorter
wavelengths. Edwards \& Snell (\cite{Edwards83}) found that FIRS\,2
was associated with a CO outflow, which further strengthened its
association with a young, intermediate PMS star. Jenness, Padman \&
Scott (\cite{Jenness95}) included FIRS\,2 in their survey of FIR cores
in the vicinity of H$_2$O masers and reported a flux density of 5.2 Jy
at 800 $\mu$m and 30 Jy at 450 $\mu$m. Eir{\'{o}}a, Palacios \& Casali
(\cite{Eiroa98}) proposed that FIRS\,2 is an intermediate mass Class~0
object, based on submillimeter photometry at 2 \,mm, 1.3 \,mm, 1.1 \,mm and
800 $\mu$m as well as from high resolution reconstructed IRAS data. 
They detected FIRS\,2 at 25, 60 and 100 $\mu$m, but not at 12
$\mu$m.  They fitted the SED of FIRS\,2 with a standard isothermal
graybody fit and found a $\beta$-index of 0.9 and a dust temperature of
35 K corresponding to a total mass of 6\Msun.

Our SCUBA data give a more accurate position for FIRS\,2 (Table~3), 
which agrees well with the H$_2$O position (uncertainty $\pm$ 10\arcsec{}). 
The source is resolved with a FWHM of $\leq$ 3.7\arcsec. The
surrounding cloud core is also seen, but because most observations were 
made using relatively short chop throws ($\sim$ 100\arcsec), the emission of
the cloud core is not well mapped. The data 
is supplemented with maps and photometry made by us using UKT14 on JCMT. Our
UKT14 flux densities are generally lower than those found by Jenness,
Padman \& Scott (\cite{Jenness95}) and Eir{\'{o}}a, Palacios and
Casali (\cite{Eiroa98}), but agree much better with the well-calibrated
SCUBA data. We find flux densities of 0.48 $\pm$ 0.10 Jy, 1.18 $\pm$
0.03 Jy, 1.69 $\pm$ 0.05 Jy and 3.66 $\pm$ 0.22 Jy for 2 mm, 1.3 mm,
1.1 mm and 800 $\mu$m, respectively. The flux densities at 2 \,mm and
1.3 \,mm are based on photometry and therefore also include a small
contribution from the surrounding cloud. The 1.1 \,mm and 800 $\mu$m
values are derived from Gaussian fits to calibrated maps, i.e.
essentially the same way SCUBA data were analyzed.

The isothermal fit (Fig. 5) predicts a $\beta$ $\sim$ 1 for
dust temperature of 42 K. The fit underestimates slightly the FIR 
data points, which are observed with a large beam size and therefore 
also include emission from the surrounding cloud core. 
Our $\beta$-index and mass (Table~4) agree rather 
well with the value derived by Eir{\'{o}}a, Palacios \& Casali 
(\cite{Eiroa98}) and demonstrate that FIRS\,2 is a cold protostellar 
source with a luminosity of $\sim$ 400\Lsun.

\section{Discussion}
\subsection{A Gas and Dust Comparison}

The $^{13}$CO emission (Fig. 1) and dust emission (Fig. 4)
show broadly similar distributions. The molecular cavity is also a dust
cavity, although the boundary appears broader in dust continuum
emission than in $^{13}$CO emission. The presence of a dust cavity
suggests that the interior is not filled with HI, but has been
substantially cleared of dust and gas. Bechis {\it et al.} (1978)
showed that radiation pressure from older stars in the cavity is
capable of doing this. Moreover, recent 21cm HI observations by H. E. Matthews et al. (2001, in preparation) in the NGC 7129 region support this hypothesis, by showing that HI in the cavity has a low number density, of 85 cm$^{-3}$, and it is in pressure equilibrium with the molecular gas in the ridge (both having P/k $\sim$ 2x10$^{5}$ cm$^{-3}$K).

In both $^{13}$CO and dust, the mapped region is seen to contain two main clumps, one surrounding the star LkH$\alpha$ 234 and extending to the northwest, the second being about 2$^{\prime}$ to the south.
Although the dust and gas maps are consistent on large scales, there
are significant differences on smaller scale. The southern clump, centered on
the peak NGC 7129 CO4, has an extension to the east with no
counterpart in the continuum maps. LkH$\alpha$ 234 SMM\,1 (identified as NGC 7129 CO1) dominates the emission in submillimeter, while NGC 7129 CO2 (SMM\,2 in Table~3) is almost equally bright in $^{13}$CO and much fainter in dust continuum.
 The peak NGC 7129 CO3 is not
present in the dust emission maps. The two $^{13}$CO peaks, NGC 7129 CO2 and CO3, have essentially equal intensity, so the difference in
continuum emission between the two is surprising. The answer may lie in
the proximity of the star SVS 13 to SMM\,2. Radiation from SVS 13 may be
responsible for maintaining the dust temperature of $\sim$ 30 K, which we
find for this source. The projected distance of SMM\,2 from SVS 13 is about three times that of NGC 7129 CO2 and therefore the stellar radiation field is less intense. However, the absence of dust emission at NGC 7129 CO3 cannot be explained by temperature alone. It must have a lower density than NGC 7129 CO2.

A quantitative comparison of gas and dust is provided by the masses
listed in Tables 2 and 5. The mass from the CO isotopomers was obtained 
from LTE expressions (Section 3.1), while the mass from the 850 $\mu$m flux
required values of $\beta$, T$_{d}$ and a gas/dust mass ratio of 100 (Hildebrand \cite{Hilde83}):

\begin{equation}
M (M_{\odot}) = 1.88 \times 10^{-{2}} \times F \times (\frac{\lambda}{0.25})^{\beta +3} \times [e^{\frac{14.4}{\lambda T_{d}}}-1] \times d^{2},
\end{equation}

\noindent
where F is the flux (in Jy) at the wavelength $\lambda$ (in mm), d
is the distance to the source (in kpc) and the mass opacity coefficient is 10 cm$^{2}$/g at 250\mic. Two values for dust masses
are given in Table~5, for different values of T and $\beta$. We will
refer below to those found with $\beta$ = 2, which is the typical value
expected from theoretical grain models (Hildebrand \cite{Hilde83}). In
view of all the assumptions that have been made, we consider the mass
values derived from the CO and from the dust to be in good agreement. In the
worst cases, the difference is a factor of three. In all these cases,
the agreement could be improved by making defensible adjustments to the
dust temperature. The results are also consistent with little or no CO 
depletion in the NGC 7129 ridge, perhaps due to the enhanced radiation field, 
which maintains grain temperatures above the 16 K CO freeze-out temperature.

\subsection{The nature of the submillimeter sources LkHa\,234\, SMM\,1 and SMM\,2}

Early studies of NGC\,7129 (Bechis {\it et al.} \cite{Bechis78}; Harvey,
Wilking \& Joy \cite{Harvey84}) indicated that the FIR emission could
be explained by dust heated by the B stars that illuminate the
reflection nebulae. The FIR emission peaked on LkH$\alpha$ 234, which
was also known to be a weak radio continuum source (Bertout \& Thum
\cite{Bertout82}), associated with an H$_2$O maser (Cesarsky {\it et al.}
\cite{Cesarsky78}) and driving a high velocity CO outflow (Edwards \&
Snell \cite{Edwards83}). The second FIR peak, FIRS\,2 was also
associated with an H$_2$O maser and a high velocity CO outflow, but
lacked an optical counterpart (Section 3.3). Since LkH$\alpha$ 234
was associated with HH objects (Hartigan \& Lada \cite{Hart85}),
coincided with a free-free emission source at 15 and 22 GHz, with
compact dust emission at 3\,mm (Wilking, Mundy \& Schwartz
\cite{Wilking86}) and drove a highly collimated optical jet (Ray {\it et al.}
\cite{Ray90}), this identification appeared rather secure.

However, Skinner, Brown \& Steward (\cite{Skinner93}) noticed that the
free-free emission was offset by $\sim$ 1\ptsec7 from the optical
position of LkH$\alpha$ 234 if they adopted the more accurate optical
position given by Herbig \& Bell (\cite{Herbig88}) or by 2\ptsec2 when 
compared with the astrometric position of Clements \& Argyle
(\cite{Clements84}). Near-IR polarimetric imaging by Weintraub, Kastner
\& Mahesh (\cite{Wein94}) showed that there is a deeply embedded
companion $\sim$ 3\arcsec\ North-West of LkH$\alpha$ 234, 
which was invisible in total intensity maps at K band. 
High resolution mid-IR imaging by Cabrit {\it et al.} (\cite{Cabrit97}) 
identified the embedded companion star,
IRS\,6, at 2\ptsec7 NW of LkH$\alpha$ 234 and showed that it has a 
steeply rising spectrum, making IRS\,6 as bright as LkH$\alpha$
234 at 17$\mu$m.  IRS\,6 therefore coincides with the radio free-free emission
source, with two H$_2$O maser spots, and with the 3\,mm dust emission
(Wilking, Mundy \& Schwartz \cite{Wilking86}; Sandell, unpublished data).  Cabrit {\it et al.} (\cite{Cabrit97}) show that IRS\,6 illuminates
an arc-shaped reflection nebula with very red colors and that it drives
an H$_2$ jet at a position angle (p.a.) of 226$^\circ$. The p.a. of the
H$_2$-jet differs from that of the optical [SII]-jet, which
has a p.a. of 252$^\circ$ (Ray {\it et al.} \cite{Ray90}). Whether the much
larger [SII] jet also originates from IRS\,6 is not clear (Cabrit et
al. \cite{Cabrit97}).  Although Mitchell \& Matthews (\cite{Mitch94})
report that the p.a. of the CO outflow agrees with the optical jet,
this is not true if the p.a. is determined from the symmetry
axis of the large scale red CO outflow, which is $\sim$
230$^\circ$. Neither does the p.a. of the optical jet agree with the
location of the two HH objects GGD\,32 and HH\,103 SW of LkH$\alpha$
234 and IRS\,6, which lie in the blue-shifted lobe of the CO outflow.

The submillimeter position for LkH$\alpha$ 234 SMM\,1 determined from our 850
$\mu$m map (Table~3) falls between LkH$\alpha$ 234 and IRS\,6, but
since our positional accuracy is $\sim$ 1\arcsec, the emission could
peak on either star. However, since the dust emission at 3\,mm peaks on
IRS\,6, it is plausible to assume that the submillimeter emission also peaks
on IRS\,6. It therefore appears that it is the cold, heavily obscured
IRS\,6 that drives the CO outflow and which is associated with the massive
dust disk or envelope we see in the submillimeter. This disk-like structure 
is orthogonal to the near-IR
H$_2$ jet to within a few degrees. From our simple isothermal modeling
(Section 3.2) we derive a luminosity of $\sim$ 9~10$^2$\Lsun, which
would make IRS\,6 an early B star, i.e.  a young Herbig Be star still
heavily enshrouded by dust. It is clear that LkH$\alpha$ 234 as well
as BD$+$65$^\circ$1638 and even BD$+$65$^\circ$1637 and SVS\,13
contribute to the 50 and 100 $\mu$m emission, but the submillimeter emission
is totally dominated by IRS\,6 and the second submillimeter source LkH$\alpha$
234\, SMM\,2. There is some dust emission in the molecular cavity as well,
especially in the vicinity of BD$+$65$^\circ$1638. This dust is hotter
than the dust in the cloud ridge, which is also seen from the spectral
index plot (Fig. 6).

The submillimeter source LkH$\alpha$ 234\, SMM\,2 does not have a known optical
or IR counterpart. It is $\sim$ 12\arcsec\ to the east of SVS\,13,
 which does not appear to be associated with any dust emission (Fig. 4). 
SMM\,2 does not coincide with the CO hotspot seen by Mitchell \& Matthews
(\cite{Mitch94}) either, which is $\sim$ 10\arcsec\ - 15\arcsec\ SE of the
submillimeter source. SMM\,2 coincides within errors with the NGC 7129 CO2 
peak (Fig. 1) and is also apparent in the CS J=3$-$2 map presented by
Fuente {\it et al.} (\cite{Fuente98}) as a dense gas
condensation.  Two of our C$^{18}$O J=3$-$2 spectra, at offsets 3 and 6 in Table~2, are close to SMM\,2, i.e. $\sim$ 5\arcsec\ north and south of the 
submillimeter peak. They show slightly elevated excitation temperatures 
(see Table~2), but the higher kinetic temperature is more likely due to 
the PDR rim than to any embedded source. The linewidths, especially in  
C$^{18}$O, are narrow and there is no evidence for outflow. SMM\,2 therefore 
has the appearance of a pre-stellar core. The virial mass is $\sim$ 8.2\Msun,
assuming a linewidth of 1 km s$^{-1}$, and a diameter of $\sim$ 14\arcsec\, 
i.e. several times larger than
the mass we derive from our submillimeter observations. However, it is clear
that the the hot gas in the cavity has compressed the whole ridge in
which SMM\,2 is embedded. We also note that our derived value for
$\beta$ as well as our assumed dust temperature is rather uncertain and
our mass estimate is at most accurate to a factor of two. SMM\,2 could therefore either be a pre-stellar core, or a protostar.

\subsection{Dust properties}

To investigate dust properties across the region, a map of the spectral index $\alpha$ ($F_{\nu} \sim \nu^{\alpha}$) was made, using the 850 and 450\mic\ deconvolved images smoothed to 14\arcsec\ resolution. Explicitly,

\begin{equation}
\alpha = \log(F_{\lambda_{1}}/F_{\lambda_{2}})/\log(\lambda_{2}/\lambda_{1}),
\end{equation}

where $\lambda_{1}$ and $\lambda_{2}$ are the two wavelengths used in our observations.

We know that the maps are distorted by chopping into the cloud emission (Section 3.2). This will affect the 450 \mic\ more than the 850 \mic\ flux, because the error lobe is larger at 450 \mic. The 450 \mic\ map was therefore corrected by adding a constant level of 5\% of the peak value to the map, before convolving it to a resolution of 14\arcsec. The resulting map is shown in Figure~6.

The spectral index map has two trends. Firstly, a comparison with the 450 $\mu$m or 850 $\mu$m map shows a general anticorrelation between submillimeter continuum flux and spectral index. The highest flux positions are minima of $\alpha$, and the value of $\alpha$ increases towards regions of weaker continuum emission. The $\alpha$ minima are seen at locations correlating with dense gas clumps, LkH$\alpha$ 234 SMM\,1, SMM\,2 and the flux peak near the offset ($-$45\ptsec0,$-$85\ptsec0). The curving ridge of 450 $\mu$m emission (Fig. 4) can be followed throughout its length in the $\alpha$ map. Secondly, the map also shows a steep rise in $\alpha$ towards the molecular cavity. This large gradient in $\alpha$ occurs along the entire transition region. 

It is important to determine whether the observed increase in $\alpha$ is due to the change in temperature through the PDR, or in the dust opacity index, $\beta$. Because F$_{\nu}$ $\propto$ $\kappa_{\nu}$$^{\beta}$B$_{\nu}$(T$_{d}$), the spectral index, $\alpha$, and the dust opacity, $\beta$, can be related as follows:

\begin{equation}
\alpha = \beta +2+\gamma,
\end{equation}

where $\gamma$ can be thought of as a correction to the Rayleigh-Jeans form of the Planck function and is given by (Visser {\it et al.} 1998)

\begin{equation}
\gamma = 1 - \log[(e^{h\nu_{850}/kT}-1)/(e^{h\nu_{450}/kT}-1)]/\log(\nu_{850}/\nu_{450}).
\end{equation}

An increase in $\alpha$ may be due to an increase in the dust temperature, T, and/or in $\beta$. The temperature is expected to be higher in the cavity because of the more intense radiation field. Our analysis in Section 3.1 showed that the gas temperature is higher towards the ridge. 
Is it possible to understand the larger values of $\alpha$ only as due to higher dust temperatures? The answer is no because $\gamma$ approaches zero at large T so that the maximum value of $\alpha$ is $\beta$ + 2. Values of $\alpha$ larger than 4 require $\beta$$\geq$2, even if T is large. While the expected increase in T$_{d}$ toward the cavity will result in higher values of $\alpha$, the increase in T$_{d}$ must be accompanied by an increase in $\beta$.

Ideally, the above relations can yield the exact value of $\beta$ if one knows the spectral index $\alpha$ and the temperature of the dust. In practice, the determination of the true value of $\beta$ is restricted by the current uncertainties in the submillimeter fluxes, especially at short wavelengths, which will reflect into uncertainties in $\alpha$. The dust will be more optically thick at 450 $\mu$m than at 850 $\mu$m (since the optical depth changes with frequency as $\tau \propto \nu^{\beta}$). Thus at 450 $\mu$m we do not see as deeply into the densest hot cores as one does at 850 $\mu$m. Another more severe restriction is our ability to accurately measure the submillimeter fluxes, especially at short wavelengths. As mentioned before, we tried to remedy this situation by adding a constant zero-level to the 450 $\mu$m flux, but this is only a first order correction. In reality, the dust structure will not be flat, but will have its own substructure, similar to that of the molecular cloud. Therefore, the numerical values of $\alpha$ in regions of low flux are highly uncertain. 

In summary, we have confidence in the values of $\alpha$ $\simeq$ 3 that we find at high flux levels. We also believe in the trend of increasing $\alpha$ towards lower flux values. In the dense cores we find $\beta$ $<$ 2, as suggested by the minima in the spectral index plot as well as our dust fits in Figure~5. Since small grains produce a high $\beta$-index and large grains lead to smaller values of $\beta$, our results are consistent with the growth of grains in regions of high density. 

These present observations of NGC 7129 support the conclusions of other similar studies that high density clumps or cores have smaller values of $\beta$ than does the surrounding gas (Visser {\it et al.} \cite{Viss98}; Johnstone $\&$ Bally \cite{Johnstone99}; Sandell {\it et al.} \cite{Sandell99}). Values of $\beta$ $\sim$ 2 outside the flux peaks have been also reported in reflection nebulae surrounding Herbig Ae/Be stars (Whitcomb {\it et al.} \cite{Whitcomb81}). Our $\alpha$ map goes further, however, in showing a general correspondence between flux and $\alpha$ over an extended area, and in showing a steep increase in $\alpha$ (and, therefore, in $\beta$) along an extended PDR. 
 
\section{SUMMARY}

New high resolution maps of the NGC 7129 ridge have been obtained, both in submillimeter continuum emission and in $^{13}$CO J=3$-$2 line emission. The main results of this study are:

1. A sharp boundary between the molecular ridge and cavity is clearly seen in the $^{13}$CO map. A previously detected narrow filament of optical and near infrared emission follows closely the cloud-cavity boundary. Strong emission in the v=1$-$0 S(1) line of H$_{2}$ from the filament, if due to fluorescence, supports the interpretation of the boundary as a PDR.

2. The $^{13}$CO emission consists of two main cloud cores connected by a bridge of gas. The northern cloud core has three $^{13}$CO peaks, while the southern one has a single extended peak.

3. The 850 $\mu$m and 450 $\mu$m maps show a very good morphological agreement with the $^{13}$CO map. Both data sets show a sharp boundary (cloud rim) toward the cavity where the two young stars BD +65 1637 and BD +65 1638 are located.

4. Three compact sources are seen in the continuum maps (SMM\,1, SMM\,2, and FIRS\,2 in Table~3). Isothermal graybody fits to the measured spectral energy distributions have been made, by combining our SCUBA data with earlier observations. The resulting masses, dust temperatures, luminosities and dust emissivities  are given in Table~4. The SMM\,1 continuum peak is identified with the deeply embedded infrared source IRS\,6, which according to these present fits is a young Herbig Be star.

5. For nine offsets around the LkH$\alpha$ 234 star, the $^{13}$CO/C$^{18}$O intensity ratio was used to obtain the mass of gas in an 14\arcsec\ beam. The masses derived from CO are found to be consistent with masses derived from the dust continuum emission. Within our analysis uncertainties the results are consistent with little or no CO depletion in this region.

6. A map of the 850$\mu$m/450$\mu$m spectral index, $\alpha$, shows a general correlation between the $\alpha$ minima and the submillimeter continuum flux peaks. The low values of $\alpha$ in the peaks also correlate with low values of dust opacity indices, $\beta$, in the range 1 to 1.6, derived for the compact source, suggesting grain growth.
The spectral index $\alpha$ rises steeply towards the molecular cavity, along the entire length of the boundary region. Consequently, the dust opacity index in this region is also high, $\beta$ $\sim$ 2. This suggests very hot, small size grains in the PDR ridge, whose mantles have been evaporated by the intense UV radiation.

\begin{acknowledgments}
 The James Clerk Maxwell
Telescope is operated on a joint basis between the United Kingdom
Particle Physics and Astronomy Research Council (PPARC), the
Netherlands Organization for the Advancement of Pure Research (ZWO),
the Canadian National Research Council (NRC), and the University of
Hawaii (UH). This research was partly supported by an operating grant from the Natural Sciences and Engineering Research Council of Canada. 
\end{acknowledgments}


\vfil\eject

\clearpage
\begin{deluxetable}{llllll}
\tablecaption{Physical parameters of the $^{13}$CO peaks. The peak antenna temperature, the line width and line velocity center were determined with a Gaussian model best fit of the spectral lines.}
\tablecolumns{6}
\tablenum{1}
\tablehead{
\colhead{Source} & \colhead{$\alpha$ (2000)} & \colhead{$\delta$ (2000)} & \colhead{${T^{*}}_{A}$} & \colhead{$\Delta$v} & \colhead{v$_{center}$}\\
\colhead{}& \colhead{($^{h}$ $^{m}$ $^{s}$)} & \colhead{($^{\circ}$ $^{\prime}$ $^{\prime\prime}$)} & \colhead{(K)} & \colhead{(km s$^{-1}$)} & \colhead{(km s$^{-1}$)} }
\startdata
NGC 7129 CO1 & 21 43 06.7 & $+$66 06 57.0 & 13.08 & 2.96 & -10.30\\
NGC 7129 CO2 & 21 43 03.5 & $+$66 07 10.0 & 15.02 & 1.21 & -10.16 \\
NGC 7129 CO3 & 21 43 02.0 & $+$66 07 28.0 & 15.89 & 1.24 & -10.12 \\
NGC 7129 CO4 & 21 42 58.5 & $+$66 05 30.0 & 17.55 & 1.44 & -10.39\\
\enddata
\end{deluxetable}

\begin{deluxetable}{llllllc}
\tablecaption{Measured properties of the gas for the nine offsets where both C$^{18}$O and $^{13}$CO observations are available\tablenotemark{*}.}
\tablecolumns{7}
\tablenum{2}
\tablehead{
\colhead{No.}&\colhead{Offset} & \colhead{${T^{*}}_{A} (^{13}CO)$} & \colhead{${T^{*}}_{A}$ (C$^{18}$O)} & \colhead{$\tau_{13}$} & \colhead{$T_{ex}$} & \colhead{$M_{gas}$} \\
\colhead{}& \colhead{(arcsec)} & \colhead{(K)} & \colhead{(K)} &\colhead{} & \colhead{(K)} & \colhead{(\Msun/14\arcsec\ beam)}}
\startdata
1 &(52.5, 37.5)    & 3.84 & 1.32  & 4.13 & 13.1 & 3.6  \\ 
2 &($-$5.0, 13.0)    & 12.57 & 4.43 & 4.27 & 29.2 & 7.3 \\ 
3 &($-$20.0, 25.0)   & 15.73 & 7.68 & 6.69 & 34.7 & 12.8 \\ 
4 &(5.0, -8.0)     & 13.13 & 3.87 & 3.35 & 30.7 & 11 \\ 
5 &(0.0, 0.0)      & 12.8 & 4.18  & 3.85 & 29.8 & 10.7 \\ 
6 &($-$22.5, 10.0)   & 14.97 & 4.96 & 3.92 & 33.6 & 7.2 \\ 
7 &($-$45.0, -85.0)  & 17.54 & 5.70 & 3.82 & 38.3 & 8.3 \\ 
8 &($-$30.0, -100.0) & 13.04 & 3.25 & 2.63 & 31.5 & 6.7 \\ 
9 &($-$20.0, -110.0) & 7.84 & 2.97  & 4.71 & 20.6 & 6.4 \\ 
\enddata
\tablenotetext{*}{The last two columns in the table were calculated assuming LTE conditions and that the gas is thermalized.}
\end{deluxetable}

\begin{deluxetable}{lllclll}
\tablecolumns{5}
\tablenum{3}
\footnotesize  
\tablewidth{450pt} 
\tablecaption{Positions and deconvolved Gaussian sizes of compact submillimeter sources.} 
\label{tbl-1}
\tablehead{
\colhead{Source} & \colhead{$\alpha$(2000.0)} & \colhead{$\delta$(2000.0)} & \colhead{$\theta_a$ $\times$ $\theta_b$} & \colhead{P.A.} \\ 
\colhead{} & \colhead{($^{h}$ $^{m}$ $^{s}$)} & \colhead{($^{\circ}$ $^{\prime}$ $^{\prime\prime}$)} & \colhead{(arcsec)} & \colhead{(degree)}  
}
\startdata
LkH$\alpha$\,234 SMM\,1  & 21 43 06.76 & $+$66 06 56.0 & 6.1 $\times$ 4.9 & $-$46 \nl
LkH$\alpha$\,234 SMM\,2  & 21 43 03.20 & $+$66 07 13.1 & 14.6 $\times$ 9.6 & $-$67 \nl
NGC\,7129\,FIRS 2 & 21 43 01.51 & $+$66 03 24.2 &  3.9 $\times$ 3.7 & n.a. \nl
\enddata
\end{deluxetable}

\begin{deluxetable}{lllllcc}
\tablecolumns{7}
\tablenum{4}  
\tablewidth{420pt} 
\tablecaption{Physical properties of the deconvolved compact submillimeter sources.} 
\label{tbl-2}
\tablehead{
\colhead{Source} & \colhead{S$_{850}$} & \colhead{S$_{450}$} & \colhead{T$_d$} & \colhead{$\beta$} & \colhead{M$_{tot}$} & \colhead{L$_{dust}$}\\ 
\colhead{} & \colhead{(Jy)} & \colhead{(Jy)} & \colhead{(K)} & \colhead{} & \colhead{(M$_{\odot})$} & \colhead{(\Lsun)}
}
\startdata
LkH$\alpha$\,234 SMM\,1    & 3.12 & 20.7 & 45 & 1.24 & 7.6 & 8.8 $\times$ 10$^2$ \nl
LkH$\alpha$\,234 SMM\,2  & 0.73 & 6.2  & 30 & 1.62 & 4.8 & 1.4 $\times$ 10$^2$ \nl
NGC\,7129\,FIRS 2    & 3.35 & 18.1 & 42 & 1.04 & 7.0 & 4.3 $\times$ 10$^2$ \nl
\enddata
\end{deluxetable}

\clearpage

\begin{deluxetable}{lccllc}
\tablecolumns{6}
\tablenum{5}
\tablewidth{460pt}
\tablecaption{Dust properties for the nine C$^{18}$O positions. The second entry for each offset uses the cloud emission only (when applicable) and a value of 2 for the $\beta$ index.}
\label{tbl-3}
\tablehead{
\colhead{Offsets} & \colhead{S$_{850\mu m}$}  & \colhead{S$_{450\mu m}$} & \colhead{T$_d$} & \colhead{$\beta$} & \colhead{M$_{tot}$}\\
\colhead{(arcsec)} &\colhead{(Jy/14\arcsec\ beam)} & \colhead{(Jy/14\arcsec\ beam)} & & \colhead{(K)} & \colhead{(M$_{\odot}$/14\arcsec\ beam)}
} 
\startdata
(52.5, 37.5)   & 0.26  & 1.39 & 13.1 & 1.71 & 6.5 \nl
               & 0.26  &      & 13.1 & 2.0  & 9.3 \nl
($-$5.0, 13.0)   & 0.74  & 6.36 & 29.2 & 1.70 & 5.5 \nl
                 & 0.43\tablenotemark{a}  &      & 25.0 & 2.0  & 8.2 \nl
($-$20.0, 25.0)  & 0.45  & 5.18 & 34.7 & 2.06 & 4.1 \nl
                 & 0.26\tablenotemark{a}  &      & 25.0 & 2.0  & 4.6 \nl
(5.0, $-$8.0)    & 1.44  & 11.6 & 30.7 & 1.58 & 8.6 \nl
                 & 0.47\tablenotemark{a}  &      & 25.0 &  2.0 & 8.5 \nl

(0.0, 0.0)     & 3.05  & 23.1 & 29.8 & 1.50 & 13.9 \nl 
               & 0.35\tablenotemark{a}  &      & 25   & 2.0  & 11.2 \nl
($-$22.5, 10.0)  & 0.47  & 4.15 & 33.6 & 1.68 & 2.8 \nl 
                 & 0.19\tablenotemark{a}  &      & 25.0 & 2.0 & 4.3 \nl
($-$45.0, $-$85.0) & 0.42  & 3.33 & 38.3 & 1.46 & 1.6 \nl
                   & 0.42  &      & 25 & 2.0 &  5.9 \nl
($-$30.0, $-$100.0)& 0.19  & 2.24 & 31.5 & 2.14 & 2.2 \nl
                   & 0.19  &      & 25.0 & 2.0  & 2.5 \nl

($-$20.0, $-$110.0)& 0.13  & 1.83 & 20.6 & 2.67 & 7.4 \nl
                   & 0.13  &      & 20.0 & 2.0     & 2.3   \nl
\enddata
\tablenotetext{a}{Cloud emission only, emission from LkH$\alpha$\,234 SMM\,1 and LkH$\alpha$\,234\, SMM\,2 has been subtracted.}
\end{deluxetable}

\clearpage


\figcaption[f1.ps]{$^{13}$CO J=3$-$2 integrated intensity map superimposed on an K$^{\prime}$ image of the region (Hodapp, 1994). The map was centered near the star LkH$\alpha$ 234, with the center coordinates $\alpha(2000)$ = 21$^{h}$ 43$^{m}$ 06$^{s}$.08 and $\delta(2000)$ = 66$^{\circ}$ 06$^{\prime}$ 56\ptsec29, and extending from $-$100\arcsec\ to +100\arcsec\ in $\alpha$ offsets and from
$-$150\arcsec\ to 75\arcsec\ in $\delta$ offsets. The crosses indicate the nine offsets with complementary C$^{18}$O J=3$-$2 data which have been chosen for analysis (the numbers correspond to Table~2 entries). \label{fig1}}

\figcaption[f2.ps]{Composite 850 $\mu$m map of LkH$\alpha$ 234 and
NGC\,7129 FIRS\,2 in grey scale overlaid with contours. The 850 $\mu$m flux 
density is contoured from 50 mJy/beam in logarithmic steps  equal to a
factor of 10$^{0.2}$ of the previous level. The  size of the beam,
14\ptsec5, is indicated at the bottom right of the plot. Near--IR sources are labeled by stars and H$_2$O masers by triangles.\label{fig2}}

\figcaption[f3.ps]{$^{13}$CO and C$^{18}$O J=3$-$2 spectral lines shown superimposed for the nine offsets marked in Figure~1 and listed in Table~2. The thin lines correspond to $^{13}$CO emission and the thick lines to C$^{18}$O emission. \label{fig3}}

\figcaption[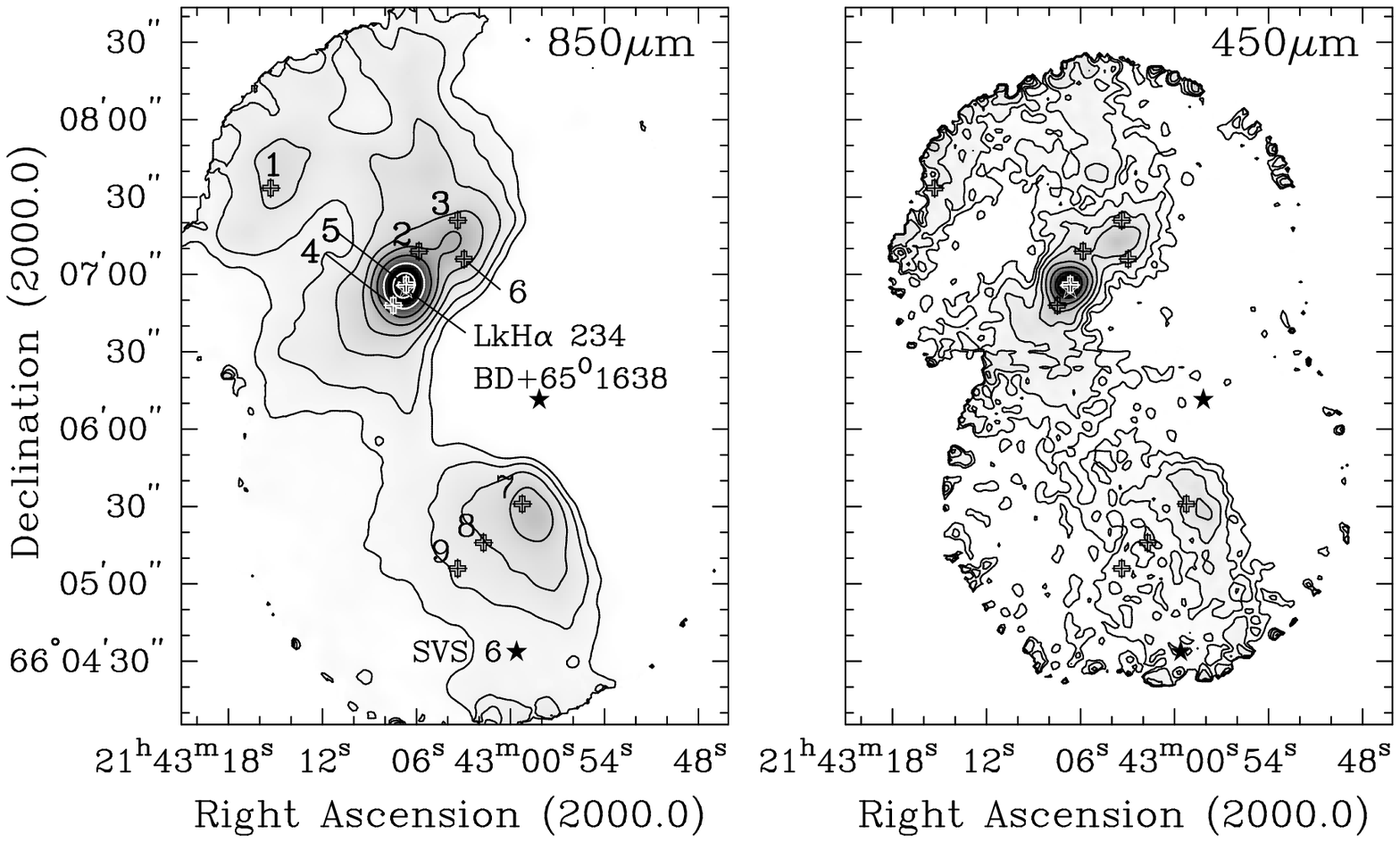]{Contour maps of deconvolved images at 850\mic\ (14\arcsec\ beam) and 450\mic\ (8\arcsec\ beam). For 850\mic\, the contour levels start from 100 mJy/beam and go in steps equal to a factor of 10$^{0.2}$ of the previous level. For 450\mic\, the contour levels start at 250 mJy/beam and go in steps of 10$^{0.25}$ of the previous level.
\label{fig4}}

\figcaption[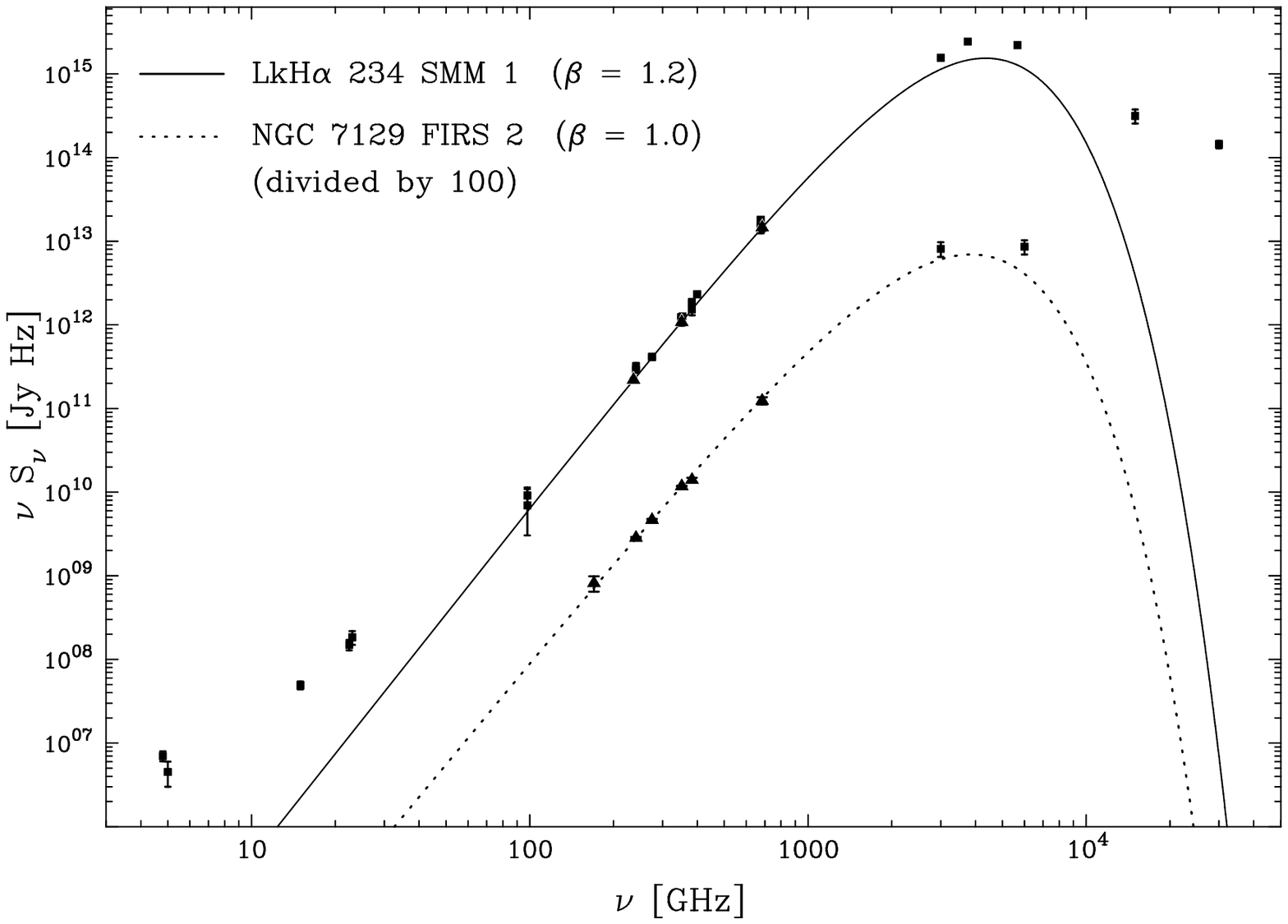]{Dust fits for LkH$\alpha$ 234 SMM\,1 and NGC7129 FIRS2, the latter data being offset by a factor of 100. The data points include our SCUBA observations, previously unpublished UKT14 data (Sandell, private communication), as well as IRAS and KAO data (Bechis {\it et al.} \cite{Bechis78}; Harvey, Wilking \& Joy \cite{Harvey84}). The latter observations were not considered in the fit, since they are likely to include hot dust from the surrounding reflection nebulosity, because of their large beam sizes. Fitted data points are marked with filled triangles, all other by filled squares.\label{fig 5}}

\figcaption[f6.ps]{Spectral index $\alpha$ map ($F_{\nu} \sim \nu^{\alpha}$) obtained from the (14\arcsec\ beamsize) 850\mic\ and 450\mic\ maps. Greyscale contours range from 3 to 5, with a step of 0.25, where dark color implies high spectral index. The optical positions are marked with stars and the LkH$\alpha$234 SMM\,2 source with a triangle.\label{fig6}} 

\end{document}